\documentclass[msmath,amssymb,aps,amsart,10pt,twocolumn]{revtex4-1}
\usepackage[utf8]{inputenc}
\usepackage{amsmath}
\usepackage{graphicx}
\usepackage{physics}
\usepackage{color,soul}
\usepackage{multirow}

\begin{document}

\title{Deviations from Poisson statistics in the spectra of free rectangular thin plates}

\author{J. L. L\'opez-Gonz\'alez}
\altaffiliation{jllopezgonzalez@correo.uaa.mx; Permanent address: Departamento de Matemáticas y Física, Universidad Autónoma de Aguascalientes}

\affiliation{Instituto de F\'isica, Universidad Aut\'onoma de San Luis Potos\'i,
 78290, San Luis Potos\'i, SLP, M\'exico.}

\author{J. A. Franco-Villafa\~ne}
\affiliation{CONACyT-Instituto de F\'isica, Universidad Aut\'onoma de San Luis Potos\'i, 78290, San Luis Potos\'i, SLP, M\'exico.}
 
\author{R. A. M\'endez-S\'anchez}
\email{mendez@icf.unam.mx}
\affiliation{Instituto de Ciencias F\'isicas, Universidad Nacional
Aut\'onoma de M\'exico, P.O. Box 48-3, 62251 Cuernavaca, Mor. M\'exico.}

\author{G. Zavala-Vivar}
\author{E. Flores-Olmedo}
\author{A. Arreola-Lucas}
\author{G. B\'aez}
\affiliation{Departamento de Ciencias B\'asicas, Universidad Aut\'onoma Metropolitana-Azcapotzalco, Ciudad de M\'exico 02200, M\'exico.}

\begin{abstract}
The counterintuitive fact that wave chaos appears in the bending spectrum of free rectangular thin plates is presented. After extensive numerical simulations, varying the ratio between the length of its sides, it is shown that (i) frequency levels belonging to different symmetry classes cross each other and (ii) for levels within the same symmetry sector, only avoided crossings appear. The consequence of anticrossings is studied by calculating the distributions of the ratio of consecutive
level spacings for each symmetry class. 
The resulting ratio distribution disagrees with the expected
Poissonian result. 
They are then compared with some well-known transition distributions between
Poisson and the Gaussian orthogonal random matrix ensemble. 
It is found that the distribution of the ratio of consecutive level spacings agrees with the prediction of the  Rosenzweig-Porter model. 
Also, the normal-mode vibration amplitudes are found experimentally on aluminum plates, before and after an avoided crossing for symmetrical-symmetrical, symmetrical-antisymmetrical, and antisymmetrical-symmetrical classes.
The measured modes show an excellent agreement with our numerical predictions. 
The expected Poissonian distribution is recovered for the simply supported rectangular plate.
\end{abstract}

\keywords{Out-of-Plane Vibrations, Rectangular Thin Plate, Avoided Crossings, 
Symmetry, Random Matrix Theory, Rosenzweig-Porter model, Free Boundary Conditions, Evanescent Waves, COMSOL, Resonant Acoustic Spectroscopy.}

\maketitle
\section{INTRODUCTION}
To understand quantum systems whose semiclassical limit is integrable or chaotic, during the last 50 years, several tools such as the theory of periodic orbits~\cite{Muller2009}, spectral statistics~\cite{Porter1965}, and the random-matrix theory (RMT)~\cite{Brody1981} have been developed. As a result of the research carried out, it was found that the main difference between wave systems with integrable or chaotic ray limit is the absence and or presence of avoided crossings also known as anticrossings. In an avoided crossing, close energy levels repel each other as a function of some externally controlled parameter. The appearance of avoided crossings can be measured as level repulsion in the spectrum, and the central paradigm of quantum chaos relies upon the association of chaotic quantum systems with level repulsion.
Those results are captured by two conjectures~\cite{Bohigas1984,Berry1977,Casati1985b}: the Berry-Tabor conjecture establishes that the spectral fluctuations of quantum systems, whose semiclassical limit is integrable, are the same as those of the Poisson distribution. The Bohigas-Giannoni-Schmit conjecture states that the spectral fluctuations of chaotic quantum systems are the same as those predicted by the Gaussian orthogonal ensemble of RMT.

The level repulsion is commonly quantified using the nearest neighbor spacing distribution $P(s)$, which measures the probability that $s$ is the distance between two consecutive levels. A simple approximate expression is the Wigner surmise $P_\mathrm{W}(s)=a_\beta~s^\beta \exp\left(- b_\beta~ s^2 - c_\beta~ s\right)$ where $a_\beta$, $b_\beta$, and $c_\beta$ are some explicitly known constants and $\beta$ is the level repulsion parameter {that holds for $\beta=0,1$}. When $\beta=0$, there is no level repulsion at all, and the distribution of the spectrum follows the Poisson law. According to the Berry-Tabor conjecture this happens in integrable systems. Chaotic systems which are invariant under time reversal, in agreement with the Bohigas-Gianonni-Schmidt conjecture, present linear repulsion ($\beta=1$). For mixed systems, intermediate values of this parameter can be found using some well-known models~\cite{Robnik2016,Izrailev1990,Batistic2013}. These results have been tested in diverse quantum, mesoscopic, and classical undulatory systems~\cite{Brody1981,Guhr1998,Stockmann1999,Dietz2015}.
Among these chaotic systems, two dimensional (2D) billiards are the most studied systems in wave chaos. In those billiards, typically, the Helmholtz equation with Dirichlet boundary conditions holds. 
The Sinai billiard~\cite{Sinai1970} and Bunimovich stadium~\cite{Bunimovich1974} are examples of chaotic systems, whereas the circle and the rectangle are integrable.
Those billiards have been extensively studied theoretically~\cite{Bogomolny1998}, numerically~\cite{Bohigas1984,Noid1980a,Noid1980b,Ramaswamy1981,Noid1983,Uzer1983}, and experimentally~\cite{Stockmann1999,Schaadt2001,Tuan2015}.

There is a plenitude of studies on vibrating plates, especially in the engineering literature~\cite{Leissa1969,Warburton1984,Soedel1993}. There are some works in a wave chaos context~\cite{Ellegaard1995,Bertelsen2000,Andersen2001,Schaadt2001,Sondergaard2002,Schaadt2003,Tanner2007}. Despite all efforts made so far, the precise nature of wave chaos in rectangular plates has remained an open question for many years. Poisson statistics are being assumed for rectangular plates without further ado.  
In this paper, the influence of the boundary conditions in the out-of-plane spectra 
of a rectangular plate is addressed. 
As it will be shown below, avoided crossings  appear  within  each  symmetry  class  of the bending spectrum for thin rectangular plates with free boundary conditions. 
Thus, deviations from the Poissonian statistics are obtained. 
To achieve that, using finite elements simulations, the level dynamics of the out-of-plane spectrum for thin rectangular plates with all its borders free is analyzed. 
The numerical results presented here extend the von Neumann-Wigner theorem~\cite{Landau1958}, for the out-of-plane vibrations of a free rectangular thin plate. Levels corresponding to different symmetry classes intersect each other, whereas this is impossible for levels within the same symmetry sector. Thus, it is shown that level repulsion appears in the frequency spectrum of the plate within each symmetry class. 
The effect of the level repulsion on the spectrum is characterized by the ratio of consecutive level spacings distribution. Since this ratio is independent of the local density of states analyzing its distribution has the advantage that unfolding, a very delicate process, is unnecessary. By testing different {heuristic} transition models between Poisson and the Gaussian orthogonal ensemble (GOE), it is shown that the Rosenzweig-Porter model best fits the numerical results. Furthermore wave amplitudes, whose eigenvalues are involved in the avoided crossing, are measured experimentally using acoustic resonant spectroscopy, a technique which has been successfully applied to study bending vibrations on integrable and chaotic plates~\cite{Manzanares-Martinez2010,Flores-Olmedo2016}. As expected, the vibrational modes exchange their identities as they pass through the avoided crossing.
%
%
\begin{figure}[tbp]
    \includegraphics[width=\columnwidth]{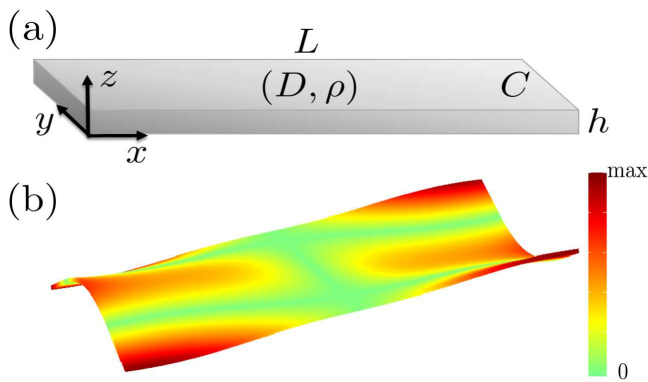} 
        \caption{(a) Rectangular plate of length $L$, width $C$ and thickness $h$; it has flexural rigidity $D$ and density $\rho$. 
        (b) Eigth bending mode at frequency $f_8=308.28$~Hz, showing its absolute vertical displacement for a simulated, free vibrating rectangular plate, where $L=800$~mm, $C=0.355$~m, $h=0.00635$~m, $E=69$~GPa, $\nu=0.33$ and $\rho=2700$~kg/m$^3$.}
       \label{Figure:Placa}
\end{figure}
\section{BENDINGWAVE SOLUTIONS FOR THE RECTANGULAR PLATE}
\begin{figure}[tbp]
    \includegraphics[width=0.8\columnwidth]{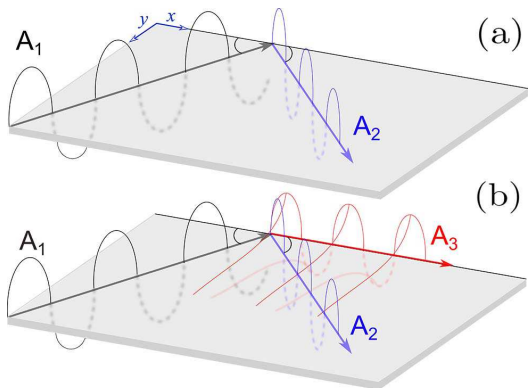} 
        \caption{Reflected bending waves on (a) a simply-supported boundary and (b) a free boundary.}
       \label{Figure:BoundaryConditions}
\end{figure}

To start with, let us consider a thin rectangular plate with length $L$, width $C$, and thickness $h$, as seen in Fig.~\ref{Figure:Placa}(a). Within the classical thin-plate theory, or Kirchhoff-Love theory~\cite{Graff1991}, the out-of-plane displacement $w(x,y,t)$ satisfies
\begin{equation}
D \nabla^4 w = -\rho h \frac{\partial^2 w }{\partial t^2},
\end{equation}
where $\nabla^4=\nabla_{\bot}^2 \nabla_{\bot}^2$ represents the biharmonic operator with 
$\nabla^2_{\bot}=\partial^2/\partial x^2+\partial^2/\partial y^2$ as the 2D Laplacian. 
Here, $D=Eh^3/12(1-\nu^2)$ is known as the plate's flexural rigidity with $E$ and $\nu$ the Young's modulus and Poisson's ratio, respectively.  
Looking for standing wave solutions $w(x,y,t)=W(x,y)e^{-i \omega t}$, one gets
\begin{equation}
\left( \nabla^4 - k^4 \right) W(x,y) =0,
\label{Equation:ThinPlate}
\end{equation}
where $k^2=\omega \sqrt{\rho h /D}$. In what follows, two types of boundary conditions for the plate will be considered. 
On the one hand, when all plate edges are free, the boundary conditions~\cite{Leissa1973} read $\left.\partial^2 W/\partial x^2+\nu\,\partial^2 W/\partial y^2\right|_{x=0,L}=0$ and $\left.\partial^3 W/\partial x^3+(2-\nu)\,\partial^3 W/\partial x\partial y^2\right|_{x=0,L}=0$. 
On the other hand, when the edges are simply supported, the boundary conditions are $\left.W\right|_{x=0,L}=0$ and $\left.\partial^2 W/\partial x^2+\nu\,\partial^2 W/\partial y^2\right|_{x=0,L}=0$. 
Interchanging $x$ and $y$, the corresponding expressions for boundary conditions on the $x$-axis at $y=0,\,C$ are obtained.
{Consider now a bending traveling wave impinging at a border of a plate, as is shown in Fig.~\ref{Figure:BoundaryConditions}. The solution is given by~\cite{Bogomolny1998,Graff1991}
\begin{equation}
\begin{split}
W(x,y)&=A_1~e^{i(k_x x - k_y y)} + A_2~e^{i(k_x x + k_y y )}\\
&+A_3~e^{-\zeta y } e^{ik_x x }
\end{split}
\label{Equation:ThinPlateSolutions}
\end{equation}
with $\zeta^2= k^2+k_x^2$ and $k_y^2= k^2-k_x^2$. For simply supported boundary conditions at the border $y=0$, just one reflected wave into the bulk of the plate  $A_2=-A_1$ appears, and the contribution from the exponential decaying term, $A_3=0$, is absent. 
In contrast for the free boundary case the exponentially decaying term is not zero, $A_3\neq 0$, generating a contribution from an evanescent wave that travels along the border as drawn in Fig.~\ref{Figure:BoundaryConditions}(b). 
This is the main difference between a free and a simply supported plate. 
} 

Due to the $D_2$ rectangle's symmetry and because the boundary conditions have the same symmetry, the solutions to Eq.~(\ref{Equation:ThinPlate}) can be classified into four symmetry classes: when modes are symmetric with respect to both $x$ and $y$-axes, they will be called symmetric-symmetric, $W^{SS}$. When the wave amplitude is antisymmetric on the $x$-axis and symmetric on the $y$-axis, $W^{AS}$, it will be called antisymmetric-symmetric. The reverse $W^{SA}$, will be called symmetric-antisymmetric. Finally, when modes are antisymmetric with respect to both axes, they will be called antisymmetric-antisymmetric $W^{AA}$.

On the one hand, for the plate with free boundary conditions, an analytical solution of Eq.~\eqref{Equation:ThinPlate} remains unknown. Then the bending spectrum and eigenfunctions have to be found numerically. 
Figure ~\ref{Figure:Placa}(b) shows a typical bending normal mode wave amplitude in which exponential (evanescent) and sinusoidal solutions can be observed. On the other hand, for simply supported boundary conditions Eq.~\eqref{Equation:ThinPlate} has normal mode analytical solutions~\cite{Leissa1973}, with frequencies given by $\omega_{nm}=(\pi/L)^2\sqrt{D/h\rho} \left[m^2+n^2 \left(L/C\right)^2\right]$. 
Here $n,m={1,2,\dots}$ label the modes; $n$ and $m$ odd (even) indicates a symmetric (antisymmetric) mode along the $x$-axis and $y$-axis, respectively. 
The corresponding simply supported bending normal mode wave amplitudes are given by $
W_{nm}=\mathcal{N}\sin\left(\sqrt{\lambda_{nm}-\pi^2 m^2}\,y/L\right) \sin\left(m\pi\,x/L \right)$, being $\mathcal{N}$ a normalization factor with $\lambda_{nm}=\omega_{nm} L^2 \sqrt{h\rho/D}$. 
Note that the wave amplitudes in the simply supported case do not present evanescent components.

\section{AVOIDED CROSSINGS IN THE FREE RECTANGULAR
PLATE AND ITS QUANTIFICATION}
\begin{figure}[tbp]
    \includegraphics[width=\columnwidth]{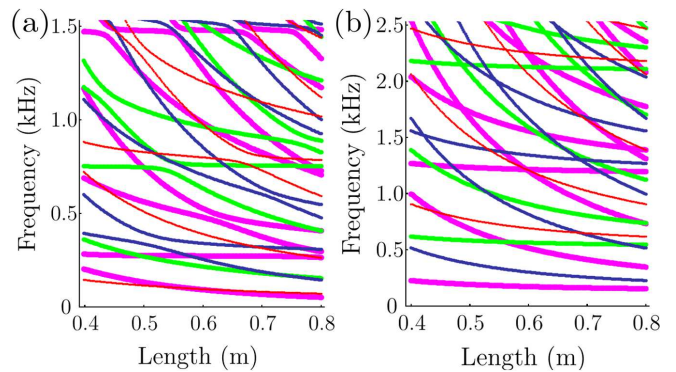} 
    \caption{
     Bending spectrum of the rectangular plate with (a) free boundary conditions and (b) with simply supported boundary conditions. The $SS$ symmetry class is drawn as thicker magenta lines, $SA$ as thick green lines, $AS$ thin blue, and $AA$ modes as thinner red lines. 
    }
    \label{Figure:EspectroTapete}
\end{figure}

Figure~\ref{Figure:EspectroTapete} shows the out-of-plane normal mode frequencies as a function of the length of the plate, keeping its width fixed. For the free plate, 70 normal mode frequencies were calculated using COMSOL Multiphysics for 200 lengths from $400$ to $800$~mm; {this software solves the 3D equations of linear elasticity, also known as Navier-Cauchy equations}. Frequencies for the $400$-mm length lie below $19$~kHz, whereas those for $800$-mm fall below $11$~kHz.
For the simply supported plate, we used the already discussed analytical normal mode frequencies. In the same figure, all symmetry classes have been distinguished by colors. In the plate with all their boundaries free, when the symmetry classes are considered independently, one can observe the presence of avoided  crossings. In contrast, only crossings appear for the simply-supported plate. {The observed avoided crossings can be understood since the free rectangular plate is not separable~\cite{Chen1991}. The appearance of evanescent waves [Eq.~\eqref{Equation:ThinPlateSolutions}] which travel along the boundary is a kind of ray splitting mechanism much slower than exponential~\cite{Couchman1992}. 
This mechanism is well known in the quantum chaos community; the new appearing orbits are called non-Newtonian orbits~\cite{Blumel1996,Sirko1997}. 
Therefore, weak avoided crossings are expected in the free rectangular plate.}

\begin{figure}[tbp]
 \includegraphics[width=\columnwidth]{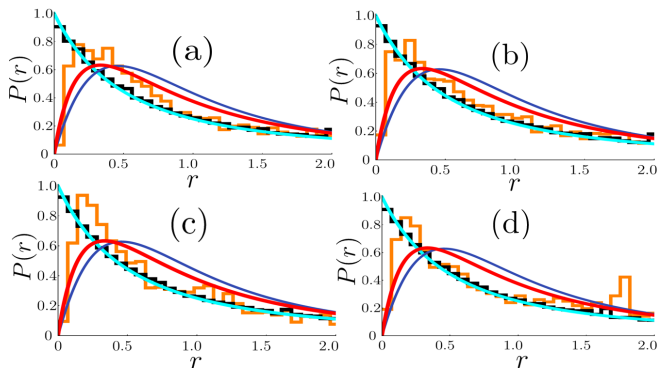}
 \caption{Profile histograms for free (orange) and simply supported (black) plate, for the ratio of consecutive level spacings within the interval $r \in [0,2]$, compared with  Poisson (cyan), GOE (violet-blue) and semi-Poisson (red)  distributions. (a) $SS$ symmetry, (b) $SA$ symmetry, (c) $AS$ symmetry and (d) $AA$ symmetry class. } 
\label{Figure:HistogramasRMT}
\end{figure}

The effect of the avoided crossings in the spectrum might be quantified using the probability distribution $P(r)$ of the ratio of two consecutive level spacings $r_n=(f_{n+1}-f_{n})/(f_{n}-f_{n-1})$. 
Here $\{f_{n}\}$ is the set of ordered normal-mode frequencies of the plate. 
The distributions of the ratio of two consecutive level spacings for the rectangular plate are shown in Fig.~\ref{Figure:HistogramasRMT} within the interval $r \in [0,2]$, and compared with the expected Poisson distribution
given by $P_{\mathrm{P}}(r)= 1/(1+r)^2$. 
Theoretical results for the Wigner surmise, a very good approximation to that of the Gaussian orthogonal ensemble,  $P_{\mathrm{W}}(r)= (27/8)(r+r^2)/(1+r+r^2)^{5/2}$ and for semi-Poisson statistics $P_{\mathrm{SP}}(r)= 6r/(1+r)^4$, are also given~\cite{Atas2013,AtasBogomolny2013}. 
As can be seen in Fig.~\ref{Figure:HistogramasRMT}, the simply supported case results agree with the Poisson ensemble's theoretical prediction; this is the expected result. 
Intriguingly, the result for the plate with free boundary conditions disagrees with the expected Poisson distribution. 
Neither the result for GOE nor semi-Poisson, seem to match the rectangular plate histograms when each symmetry class is independently considered. Then the free bending vibrations of the rectangular plate are neither fully integrable nor chaotic. 
Also from Fig.~\ref{Figure:HistogramasRMT}, it is clear that the Poisson distribution does not correctly predict the free bending histograms in particular for $r<0.2$. 
The difference in this region suggests that the chaotic contribution, coming from the avoided crossings, is small but significant for $r\rightarrow0$.

\section{RMT TRANSITION MODELS FITS FOR THE
DISTRIBUTION OF THE RATIO BETWEEN
CONSECUTIVE LEVEL SPACINGS}
The behavior of the distribution of the ratio between consecutive level spacings for the rectangular plate with free boundary conditions opens up the possibility of exploring the agreement with transition models between Poisson and GOE. 
Although there are many {heuristic} models to explain such a transition~\cite{Robnik2016,Seligman1984,Berry_1984}, only three models will be considered here.
Let us first take the phenomenological attempt given by Brody-Atas (BA)~\cite{Atas2013,Brody1973}
\begin{equation}
P_\mathrm{BA}(r,\beta)= \frac{1}{Z_{\beta}}\frac{(r+r^2)^{\beta}}{(1+r+r^2)^{1+(3/2)\beta}},
\label{Equation:PrBrodyAtas}
\end{equation}
where $Z_\beta$ is obtained from the normalization; $Z_1=8/27$ for GOE ($\beta=1$). As mentioned before, $\beta$ indicates the level repulsion parameter. Another transition model considered is the Rosenzweig-Porter (RP) model initially built to adjust nuclear spectra halfway between Poisson and GOE distributions~\cite{Rosenzweig1960}.
The RP model considers that the physical system under study has a Hamiltonian $ H_{\lambda}$ that shows an integrable behavior (Poisson) plus a chaotic behavior (GOE) through the variation of a continuous parameter $\lambda$~\cite{Leyvraz1990,Lenz1991}: 
\begin{equation}
H_{\lambda}= \frac{H_0 + \lambda V}{\sqrt{1+\lambda^2}},
\label{Equation:PMLG}
\end{equation}
where $H_0$ represents a diagonal matrix with $d$(=10000)
 independent Gaussian variables centered on zero and variance equal to 1; $V$ represents a GOE  
matrix with independent Gaussian variables centered at zero and variance equal to $\sigma^2$ (except the diagonal elements where its variance is $2\sigma^2$).
For $\lambda = 0$ this model reveals Poisson type statistics and for $\lambda \rightarrow \infty$ it shows GOE statistics~\cite{Chavda2014}.
The last distribution considered in this work is $P_\mathrm{I}(r)$ based on the Izrailev transition model~\cite{Izrailev1990} {that generalizes the Wigner surmise for any value of $\beta$ between 0 and 1}.  The procedure to build up $P_{I}(r)$ from $P_{\beta}(s)$ is described as follows. A set of spacings $\{ s_i\}$ is created from the Izrailev's distribution
\begin{equation}
P_{\beta}(s)=  A\left( \frac{1}{2}\pi s \right)^{\beta} 
\! \exp \left[-\frac{\beta}{16}  \pi^2 s^2 - \left( B - \frac{\pi}{4}  \beta \right) s \right],
\label{Equation:Izrailev}
\end{equation}
with parameters $A$ and $B$ found from the conditions $\int_{0}^{\infty} P_{\beta}(s)\hspace{0.09cm} ds = 1$ and $\int_{0}^{\infty} s \hspace{0.09cm} P_{\beta}(s)\hspace{0.09cm} ds = 1$.
Then, a set of ratios $\{r_i\}$ is built and its distribution $P(r)$ is obtained. Note that the previous procedure can be applied to any transition model given its nearest-neighbor spacing distribution.

\begin{figure}[tbp]
 \includegraphics[width=\columnwidth]{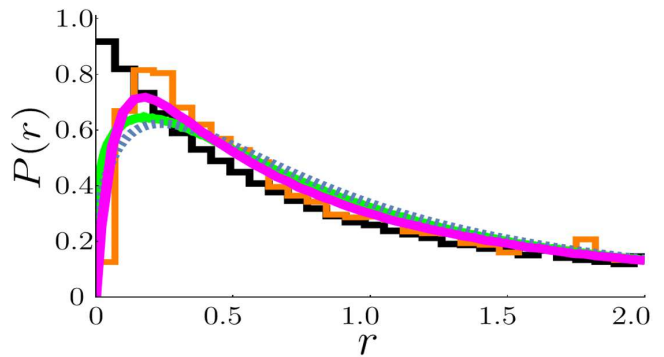}
 \caption{Average histogram profile for all symmetry classes of the free plate (orange) compared with the best fit of RMT transition models: Brody-Atas (dotted blue), Izrailev (green) and Rosenzweig-Porter (magenta); also the average profile histogram distribution for the simply supported plate is shown (black).}
\label{Figure:4}
\end{figure}

The best fits obtained for the BA, RP, and Izrailev models are shown in Fig.~\ref{Figure:4}. 
For BA and Izrailev ensembles, we fitted the level repulsion parameter $\beta$.
For the RP model, the parameter $\lambda$ was fitted. {Table~\ref{Tabla:1} contains the comparative results for all fitted distributions using Pearson's chi-squared ($\chi^2$) test.} 
The mean absolute percentage error within the interval $r \in [0,2]$ for the models in descending order are $20.3$~\%, $18.6$~\% and $12.1$~\% for BA ($\beta=0.3459$), Izrailev ($\beta=0.25$), and RP ($\lambda=5.85\times10^{-3}$), respectively. 
The RP model with a small $\lambda$ shows a better agreement. 
The BA and Izrailev best fits report almost the same repulsion parameter around $\beta\sim 0.3$. 
{It is well known that the value of $\beta$ depends on the missing levels~\cite{Molina2007, Bialous2016, Dietz2017} which are not considered in the statistics. In our case, by checking in detail the out-of-plane modes we found very few missing levels, representing at most $0.26$~\% of any symmetry class. Therefore, this small number of missing levels does not affect the reported $\beta$.}
The GOE and Poisson distributions (not shown in Fig.~\ref{Figure:4}) have even bigger errors, $36.6$~\% and $30.9$~\% respectively. 
\begin {table}[ht!]
\caption{Fit comparison of RMT models and the average histogram over all symmetries, for the free rectangular thin plate. Number of bins $=30$.}
\begin{center}
\begin{tabular}{ |p{2cm}||p{2.2cm}|p{1.5cm}|p{1.7cm}|  }
\hline
Distribution& Parameters& Error (\%)& $\chi^2$/(bins-1)\\
 \hline
 Poisson &  $\cdots$  & 30.93 & 0.032\\
  Semi-Poisson& $\cdots$  & 24.37 & 0.023\\
 GOE & $\cdots$ & 36.65 & 0.058\\
 Brody-Atas & $\beta$=0.3459 & 20.30& 0.016\\
  Izrailev &  $\beta$=0.25 & 18.68& 0.015\\
 Rosenzweig-Porter
 & $\lambda$=0.00585  & 12.17& 0.008\\
 \hline
\end{tabular}
\end{center}
\label{Tabla:1}
\end{table}
\section{WAVE AMPLITUDE MEASUREMENTS THROUGH
AVOIDED CROSSINGS}
Avoided crossings are the fundamental and differentiating effect that appears in the free-bending vibrations of the rectangular plate. 
Thus, in what follows, measurements that prove the existence of avoided crossings for three of the four symmetry classes are reported. 
A generalization of the resonant acoustic spectroscopy technique~\cite{Flores-Olmedo2016,Arreola-Lucas2015} was used to verify the shape of the predicted stationary patterns ``before'' and ``after'' an avoided crossing. 
Three aluminum plates with identical mechanical properties were used in the experiments, with a width of $355$~mm, thickness of $6.35$~mm, and lengths of $400$, $500$, and $800$~mm.
These lengths were selected considering the numerical predictions shown in Fig.~\ref{Figure:EspectroTapete} in order to observe the avoided crossings within each symmetry class. {To rest the plate, nylon threads in cross shape were held. The weight of the plate tensions the threads. This assembly allows almost free-boundary vibration of the plate that is weakly disturbed in four points by the threads. More details can be found in Refs.~\cite{Flores-Olmedo2016,Arreola-Lucas2015}.
\begin{figure}[tbp]
 \includegraphics[width=\columnwidth]{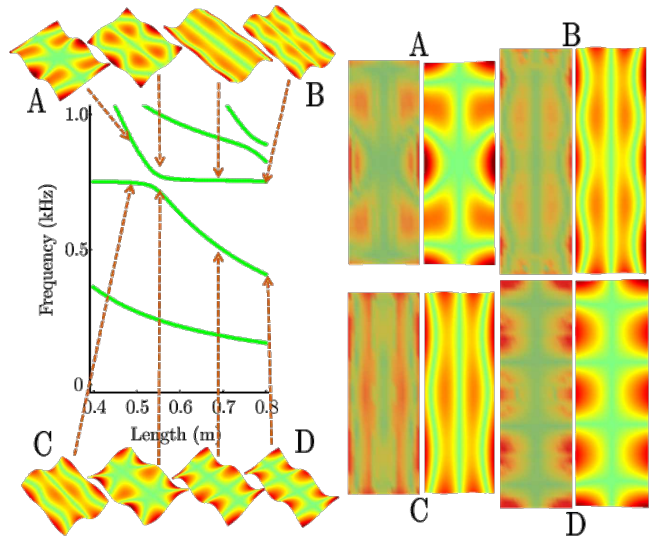}
\caption{Left: the frequency spectrum as a function of the length $L$ for the SA symmetry pointing a sequence of bending wave amplitudes numerically calculated, through an avoided crossing, for the free rectangular plate. The colors follow the same scale as in Fig.~\ref{Figure:Placa}, indicating the nodes by green and maxima by red. Right: comparison between measured modes (darker) and their corresponding simulated ones (brighter). Modes A and C correspond to a $500$~mm length plate, while modes B and D to an $800$~mm length plate. 
}
\label{Figure:5}
\end{figure}

The left-hand side of Fig.~\ref{Figure:5} shows the sequence of bending wave amplitudes obtained numerically, through an avoided crossing, for the SA symmetry of the plate. Colors for these stationary patterns obey the color scale of Fig.~\ref{Figure:Placa}(b), indicating the nodes by green and the maxima or minima by red. 
The upper sequence illustrates the evolution from a stationary 2D pattern A to a quasi-1D pattern B. Pattern A has four nodal lines (green color) in one direction and one nodal line in the perpendicular direction, while pattern B has three nodal lines along this last direction. The lower sequence shows in turn an inverse evolution, that is, the amplitude C evolves until it becomes pattern D. Note that B and C are practically the same patterns, as well as A and D.
At the right-hand side of Fig.~\ref{Figure:5}, a comparison between numerical calculations and experimental realizations for each one of the four stationary patterns associated with the avoided crossing is made. Each image is an $x-y$ plane projection of the absolute vertical out-of-plane wave amplitude. The smooth and continuous images (second and fourth columns) correspond to simulation, in contrast to experimental patterns (first and third columns) whose mapping is less intense and defined. The agreement between experiment and numerics is remarkable for all patterns. The error in frequency between the experimental and numerical predictions was at most $3.2$~\% for all measured modes within each symmetry sector.
\begin{figure}[tbp]
 \includegraphics[width=\columnwidth]{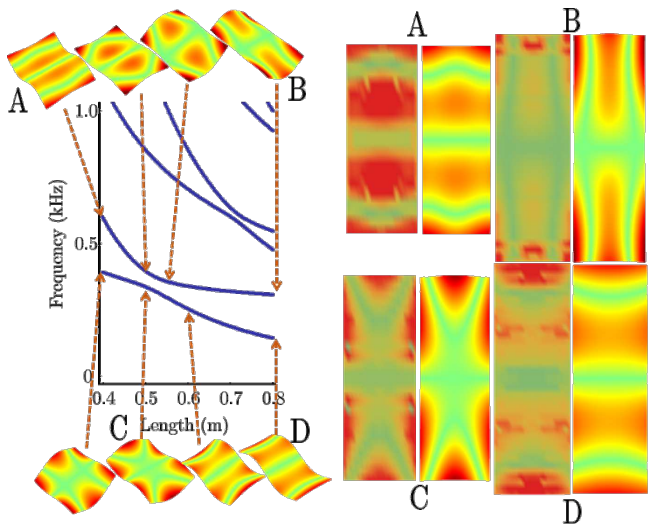}
\caption{Evolution through an avoided crossing for the AS symmetry class. The figure is ordered as in Fig.~\ref{Figure:5}. Mode A is for a plate with $L=400$~mm and mode C for $L=500$~mm; modes B and D correspond to $L=800$~mm.
}
\label{Figure:6}
\end{figure}

In Fig.~\ref{Figure:6} a comparison of stationary patterns for the AS symmetry through an avoided crossing, similar to that of Fig.~\ref{Figure:5}, is given.
As can be seen, the interchange of both patterns before and after the anticrossing is observed. For this symmetry a 2D normal mode indicated by C was chosen to be measured. A plate with $L=400$~mm was used since it was difficult to distinguish experimentally the upper mode A before the avoided crossing for $L=500$~mm. 
The comparison between experimental stationary patterns A, B, C, and D versus the numerical ones shows an excellent agreement and one-to-one correspondence.
\begin{figure}[tbp]
 \includegraphics[width=\columnwidth]{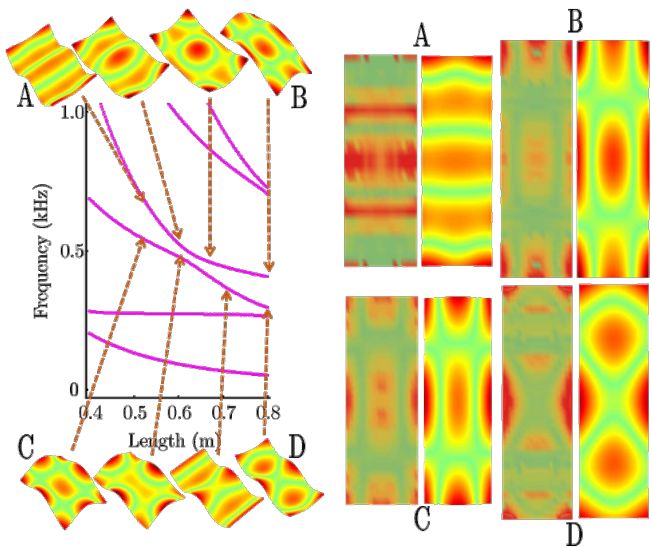}
\caption{Evolution through an avoided crossing for the SS symmetry class. The figure is ordered as in Fig.~\ref{Figure:5}. Modes A and C correspond to a plate with $L=500$~mm; modes B and D are for $L=800$~mm.
}
\label{Figure:7}
\end{figure}

Figure~\ref{Figure:7} shows the evolution across an avoided crossing for the SS symmetry. An apparent discrepancy is observed between equivalent numerical patterns A and D before and after the anticrossing. Pattern D was chosen to be measured experimentally but as we can see from the lower branch evolution, a plate with $L=700$~mm would be more appropriate to check the similarity to pattern A. More experiments around this length have to be performed to get a better agreement. 
In general, the comparison between numerical predictions and experiment reveals good agreement.}
\section{CONCLUSIONS}
The generic presence of avoided crossings in the bending spectrum of freely vibrating rectangular thin plates has been reported.
This was done numerically up to $20$~kHz and verified experimentally for low frequencies. 
The avoided crossings are responsible for non-Poissonian behavior in the spectrum statistics characterized by the ratio of consecutive level spacings distribution. 
In the rectangular free plate, the Berry-Tabor conjecture does not apply due to the following reason.
When a bending ray arrives at a free boundary, apart from the reflected ray, a new ray that travels only along the boundary appears. This ray splitting implies weak avoided crossings.
The Poissonian statistics are recovered for simply supported boundary conditions since evanescent waves on the boundary are not present. 
Several RMT transition models were tested, and the one that best fits the spectrum statistics of the free plate is the RP model. The present work opens the door to a plethora of developments in wave chaos since only two of 21 different boundary conditions of a rectangular vibrating plate have been analyzed. 
\begin{acknowledgments}
This work was initiated as an academic interchange program between CIC-UNAM and UAA; it was supported by DGAPA-UNAM under Project No. PAPIIT IN109318. R.A.M.S., G.B. and J.A.F.V. received funding from CONACYT under Projects No. 284096, No. 285776, No. A1-S-33920, and No. A1-S-18696, respectively. J.L.L.G. received financial support by CONACYT, FAI-UASLP, INVESTEL and IDSCEA. The authors collaborated under the ``Waves and Metamaterials Group” initiative. We thank R. Diamant, T. Seligman, F. Leyvraz, and A. Martínez-Argüello for fruitful discussions.
We also appreciate Centro Internacional de Ciencias A.C. for opening its facilities to group meetings and gatherings hosted there.
\end{acknowledgments}

\nocite{*}
\bibliography{BiblioRectangularPlate}

\end{document}